\documentclass[12pt]{article}
\usepackage{graphicx}
\usepackage{amsmath}

\newcommand{\NDA}{\Omega_{\rm NDA}}
\newcommand{\gtrsim}{\mathop{>}\limits_{\displaystyle{\sim}}}

\makeatletter
\def\@maketitle{\newpage
 \null
 {\normalsize \tt \begin{flushright} 
  \begin{tabular}[t]{l} \@date 
  \end{tabular}
 \end{flushright}}
 \begin{center}
 \vskip 2em
 {\LARGE \@title \par} \vskip 2.5em {\large \lineskip .5em
 \begin{tabular}[t]{c}\@author 
 \end{tabular}\par} 
 \end{center}
 \par
 \vskip 1.5em} 
\makeatother
\title{Dynamical Electroweak Symmetry Breaking from Extra Dimensions\footnote{
        \uppercase{T}alk given by \uppercase{M}.\uppercase{H}ashimoto.}}
\date{DPNU-03-07 \\ TU-686 \\ UT-ICEPP 03-03 \\ March 2003}
\author{
  {\large Michio {\sc Hashimoto}}\\[1mm]
  {\it \normalsize ICEPP, the University of Tokyo, }\\[-1mm]
  {\it \normalsize Hongo 7-3-1, Bunkyo-ku, }\\[-1mm]
  {\it \normalsize Tokyo 113-0033, Japan}\\[-1mm]
  {\tt \normalsize E-mail: michioh@post.kek.jp}\\[7mm]
  {\large Masaharu {\sc Tanabashi}}\\[1mm]
  {\it \normalsize Department of Physics, Tohoku University}\\[-1mm]
  {\it \normalsize Sendai 980-8578, Japan}\\[-1mm]
  {\tt \normalsize E-mail: tanabash@tuhep.phys.tohoku.ac.jp}\\[7mm]
  {\large Koichi {\sc Yamawaki}}\\[1mm]
  {\it \normalsize Department of Physics, Nagoya University} \\[-1mm]
  {\it \normalsize Nagoya 464-8602, Japan}\\[-1mm]
  {\tt \normalsize E-mail: yamawaki@eken.phys.nagoya-u.ac.jp}
}

\begin{document}
\maketitle

\vspace*{1.0cm}

\begin{center}
  Presented at \\
  {\it 2002 International Workshop On}\\
  {\it "Strong Coupling Gauge Theories}\\
  {\it and Effective Field Theories" (SCGT 02)} \\
  {\it Nagoya, Japan} \\
  {\it 10--13 December 2002}
\end{center}

\newpage

\begin{abstract}
We study the dynamical electroweak symmetry breaking (DEWSB) in the 
$D (=6,8,\cdots)$-dimensional bulk with compactified extra dimensions.
We identify the critical binding strength for triggering the DEWSB,
based on the ladder Schwinger-Dyson equation.
In the top mode standard model with extra dimensions,
where the standard model gauge bosons and the third generation of
quarks and leptons are put in 
the bulk,
we analyze the most attractive channel (MAC) 
by using renormalization group equations (RGEs) of 
(dimensionless) bulk gauge couplings
and determine the effective cutoff where the MAC coupling 
exceeds the critical value.
We then find that 
the top-condensation can take place for $D=8$. 
Combining RGEs of top-Yukawa and Higgs-quartic couplings with 
compositeness conditions, 
we predict 
the top mass, $m_t=173-180$ GeV, and the Higgs mass,
$m_H=181-211$ GeV, for $D=8$,
where we took the universal compactification scale $1/R = 1-100$ TeV.
\end{abstract}

\section{Introduction}
The  origin of mass, or  
the electroweak symmetry breaking (EWSB)
is one of the most urgent problems of the particle physics today.
There are some dynamical 
approaches 
toward this problem such as Technicolor~\cite{Weinberg},
top-quark condensate~\cite{MTY89,Nambu89}, etc.~\cite{Hill:2002ap}.
In the top-quark condensate
or the ``top mode standard model'' (TMSM),
the scalar bound state of $\bar{t}t$ plays the role of 
the Higgs boson in the SM and 
the top quark naturally acquires the dynamical mass of the order of 
the EWSB scale.
However, the original version of the TMSM~\cite{MTY89}  has some problems:
The 4-fermion interaction of the top-quark
was introduced by hand
in order to trigger the EWSB.
The top-quark mass $m_t$ 
was predicted
somewhat larger than the experimental value, $m_t \gtrsim 200 {\rm GeV}$, 
even if we take the cutoff to the Planck or the GUT 
scale.~\cite{MTY89,Bardeen:1989ds,Hashimoto:1998tj} 
Such a huge cutoff also causes a serious fine-tuning problem. 

Recently, Arkani-Hamed, Cheng, Dobrescu, and Hall 
(ACDH)~\cite{Arkani-Hamed:2000hv} proposed 
an interesting version of TMSM
in extra dimensions, where
the SM gauge bosons and some generations of quarks and leptons
are embedded in $D(=6,8,\cdots)$ dimensions.
Since bulk gauge interactions become naively non-perturbative
in the high-energy region,
the bulk QCD may trigger the top-condensation without
introducing ad hoc 4-fermion interactions.
Moreover, all condensates of Kaluza-Klein (KK) modes of the top quark
contribute to the EWSB,
thereby suppressing the predicted value of $m_t$.

However, we have found that the bulk QCD coupling 
has an upper bound.~\cite{Hashimoto:2000uk}
It is thus nontrivial whether the EWSB dynamically takes place 
due to the bulk QCD or not.
We thus studied the dynamics of bulk gauge theories,
based on the ladder Schwinger-Dyson (SD) 
equation.~\cite{Hashimoto:2000uk,Gusynin:2002cu}
By this we estimated the critical value of the binding strength
$\kappa_D^{\rm crit}$, which particularly disfavored the 
top-quark condensate in the $D=6$ case of the  
ACDH version. (See Sec.~2.) 

Now, in order for the ACDH version to work phenomenologically, 
the top-condensate should be the most attractive channel 
(MAC)~\cite{Raby:1979my} 
and its binding strength $\kappa_t$ at the cutoff $\Lambda$ 
should exceed $\kappa_D^{\rm crit}$, 
whereas those of other bound states such as the tau-condensate 
($\kappa_\tau$) should not:
\begin{equation}
  \kappa_t(\Lambda) > \kappa_D^{\rm crit} > 
  \kappa_\tau(\Lambda), \cdots. \label{top-cond}
\end{equation}
Comparing our estimations of $\kappa_D^{\rm crit}$ with the 
binding strengths obeying the
renormalization group equations (RGEs) of the gauge couplings,
we determine the cutoff $\Lambda$ satisfying Eq.~(\ref{top-cond}) and 
thereby predict the top mass as well as the Higgs mass.~\cite{future}
In fact we can obtain a certain region of the effective cutoff 
$\Lambda$ satisfying Eq.~(\ref{top-cond}) for $D=8$,
while the top-condensation is unlikely to occur for
$D=6$. (See Sec.~3.)
This is in sharp contrast to the earlier approaches 
of ACDH~\cite{Arkani-Hamed:2000hv} and Kobakhidze~\cite{Kobakhidze:1999ce} 
where the cutoff $\Lambda$ is treated as an adjustable parameter.

In Sec.~4, we predict masses of the top quark and 
the Higgs boson in the formulation  \'{a} la Bardeen, Hill, 
and Lindner (BHL)~\cite{Bardeen:1989ds} based on 
RGEs and compositeness conditions.
The value of $m_t$ around the universal compactification scale
$1/R$ is governed by the quasi infrared-fixed point (IR-FP) 
for the top-Yukawa coupling $y_*$~\cite{Hill:IR-FP}.
The behavior of $y_*$ is approximately given by 
$y_*^2/g_3^2 = C_F (6+\delta)/(2^{\delta/2}N_c + 3/2)$, 
where $C_F=(N_c^2-1)/(2N_c)$, $N_c=3$, $\delta \equiv D-4$, and 
$g_3$ is the conventional QCD coupling. 
Thus, the prediction of $m_t$ can be suppressed as $\delta$ increases.
We predict numerically the top mass, $m_t = 173-180$ GeV, and 
the Higgs mass, $m_H=181-211$ GeV, for $D=8$,
where we took $1/R=1-100$ TeV.

\section{Analysis of the ladder SD equation}

We consider that the SM gauge group and the third generation
of quarks and leptons are put in the $D(=6,8,\cdots)$-dimensional bulk,
while the first and second generations are confined in 3-brane (4-dimensions).
We assume that four of $D$-dimensions are 
the usual Minkowski spacetime and extra $\delta$ spatial dimensions
are compactified at a universal scale $1/R \sim {\mathcal O}(\rm TeV)$.

Before analyzing the ladder SD equation,
we study running effects of dimensionless bulk gauge couplings $\hat g_i$
$(i=3,2,Y$). 
Above the compactification scale $1/R$, we should take into account 
contributions of Kaluza-Klein (KK) modes.
We find approximately the total number of KK modes $N_{\rm KK}(\mu)$ 
below the renormalization point $\mu$,
\begin{equation}
  N_{\rm KK}(\mu) =
  \frac{1}{2^n}\frac{\pi^{\delta/2}}{\Gamma(1+\delta/2)}(\mu R)^\delta,
  \quad (\mu \gg 1/R) \label{n_kk}
\end{equation}
with the orbifold compactification $T^{\delta}/Z_2^n$,
where we take $Z_2$ and $Z_2 \times Z'_2$ projections for $D=6$ and 
$D=8$, respectively.
The dimensionfull bulk gauge coupling $g_D$ is related to 
the conventional 4-dimensional gauge coupling $g$ as
$g_D^2 = g^2 \cdot (2\pi R)^\delta/2^n$.
Combining the definition of $\hat g (\equiv g_D \mu^{\delta/2})$ with
RGEs for $g_i$ and Eq.~(\ref{n_kk}),
we obtain approximately RGEs for $\hat g_i$, 
\begin{equation}
 \mu \frac{d}{d \mu} \hat g_i = \frac{\delta}{2}\hat g_i
 + (1+\delta/2) \NDA b'_i\, \hat g_i^3 , 
 \quad (\mu \gg 1/R) \label{rge_ED3}
\end{equation}
with $\NDA \equiv [(4\pi)^{D/2}\Gamma(D/2)]^{-1}$.
RGE coefficients $b'_3$ and $b'_Y$ are given by
\begin{equation}
  b'_3 = -11+\frac{\delta}{2}+\frac{4}{3}\cdot 2^{\delta/2} \cdot n_g,
  \qquad 
  b'_Y = \frac{20}{9}\cdot 2^{\delta/2} \cdot n_g + \frac{1}{6}n_H ,
\end{equation}
where $n_g$ $(n_H)$ denotes
number of generations (composite Higgs bosons) in the bulk.
Hereafter, we assume $n_g=1,n_H=1$.
We note $b'_3 < 0$ for $D=6,n_g=1,2,3$ and $D=8,n_g=1$.
We find that the bulk QCD coupling with $b'_3 < 0$ has 
the ultraviolet fixed-point (UV-FP) $g_{3*}$,
\begin{equation}
 g_{3*}^2 \NDA = \frac{1}{-(1+2/\delta)\,b'_3}
\end{equation}
by using Eq.~(\ref{rge_ED3}).
We can also show that
the UV-FP is the upper bound of $\hat g_3^2$.
We thus need to investigate whether the bulk QCD coupling can be 
sufficiently large to trigger the EWSB or not.

\begin{figure}[tbp]
  \begin{center}
  \resizebox{0.7\textwidth}{!}
            {\includegraphics{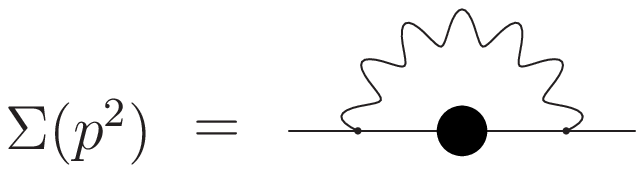}}
  \caption{The ladder SD equation. The solid and wavy lines 
           with and without the shaded blob represent 
           the full propagator of the fermion and the bare
           one of the gauge boson, respectively. 
           The mass function of the fermion is
           written as $\Sigma(p^2)$ with the external (Euclidean) momentum $p$.
           \label{fig:sd}}
  \end{center}
\end{figure}

Let us investigate the condition for the EWSB triggered by 
the bulk QCD.
The $D$-dimensional ladder SD equation
is given by Fig.~\ref{fig:sd}:
\begin{equation}
  \Sigma(p^2)=\int \frac{d^D q}{(2\pi)^D}\frac{\Sigma(q^2)}{q^2+\Sigma(q^2)}
  \frac{(D-1) C_F g_D^2}{(p-q)^2} \label{sd-eq} 
\end{equation}
with the mass function $\Sigma$ and Euclidean momenta $p,q$, 
where we took the Landau gauge in order to make 
the wave function renormalization identically unity.~\cite{Hashimoto:2000uk}
For simplicity, we incorporate running effects of the bulk QCD
as $\kappa_D (\mu) \equiv C_F \hat g_3^2 (\mu) \NDA ={\rm const.}$,
which is closely related to the MAC coupling, just on the UV-FP. 
This simplification obviously makes the critical point lower.
We then obtain the critical binding strength $\kappa_D^{\rm crit}$:
\begin{equation}
 \kappa_6^{\rm crit} \simeq 0.122, \quad  \kappa_8^{\rm crit} \simeq 0.146
 \label{kd_sd1}
\end{equation}
for $D=6$ and $D=8$, respectively.~\cite{Hashimoto:2000uk}
These are minimal values within uncertainties of the ladder SD 
equation.~\cite{Hashimoto:2000uk,Gusynin:2002cu}
The critical points are unlikely to be smaller than the above values, 
even if we take into account the effect that the cutoff $\Lambda$ is not 
so large in fact as compared with the compactification scale
$1/R$.~\cite{Hashimoto:2000uk,Gusynin:2002cu}
We use most conservatively the values of Eq.~(\ref{kd_sd1}) 
in the following analysis. 

The upper bounds of the bulk QCD coupling in the ACDH scenario are found
as $\kappa_{6,8}=0.091,0.242$ for $D=6,8$.~\cite{Hashimoto:2000uk}
Thus, the top-condensation is unlikely to occur for $D=6$.

\section{Conditions for the top-condensation in the bulk}

We analyze the MAC at the cutoff $\Lambda$ by using RGEs of bulk gauge 
couplings.
In the one-gauge-boson-exchange approximation, 
we can easily obtain $\kappa_{t,\tau}$: 
\begin{equation}
  \kappa_t (\mu)= C_F \hat g_3^2 (\mu) \NDA 
                 + \frac{1}{9}\hat g_Y^2 (\mu) \NDA ,
\end{equation}
\begin{equation}
  \kappa_\tau (\mu) = \frac{1}{2} \hat g_Y^2 (\mu) \NDA . 
\end{equation}
The MAC is the top (tau)-condensate, 
when the bulk QCD (hypercharge) is dominant.

\begin{figure}[tbp]
  \begin{center}
  \resizebox{0.47\textwidth}{!}
            {\includegraphics{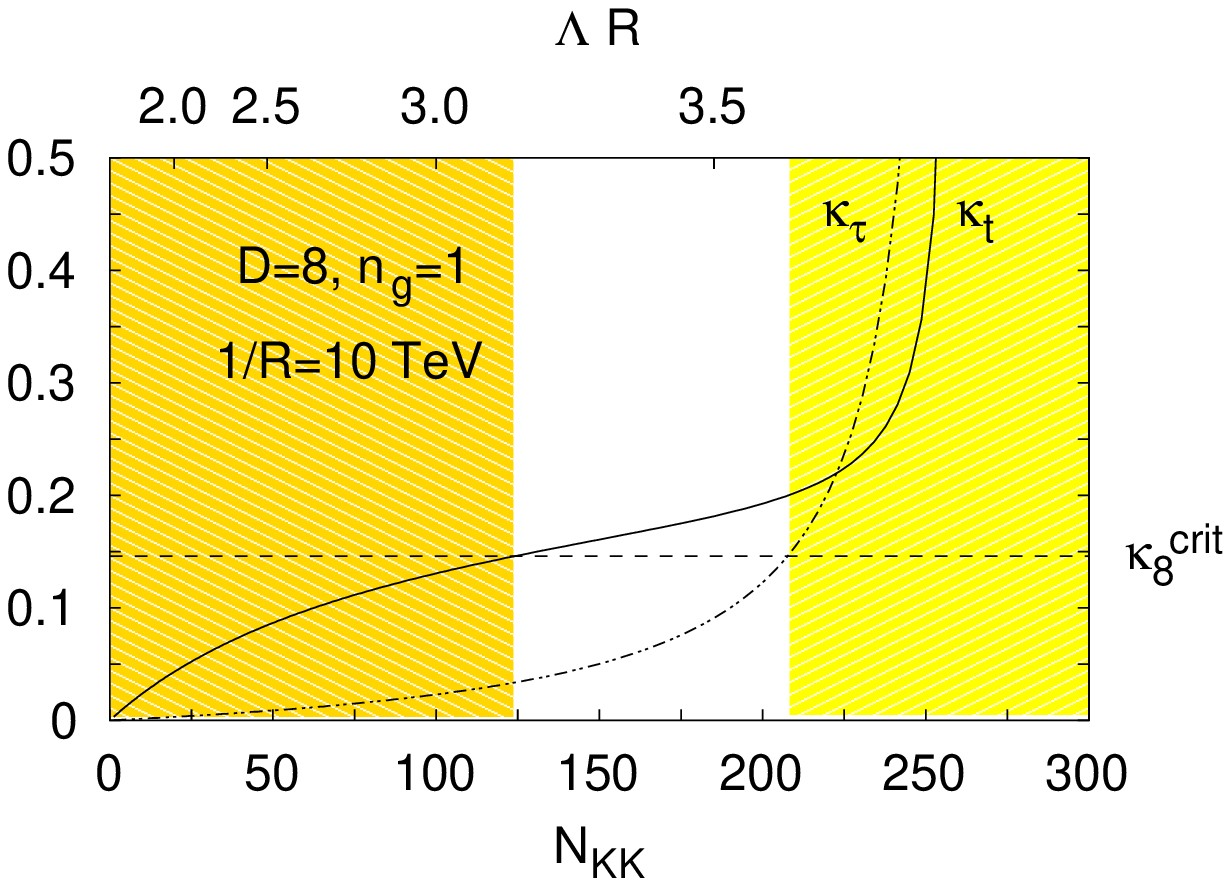}} \quad 
  \resizebox{0.47\textwidth}{!}
            {\includegraphics{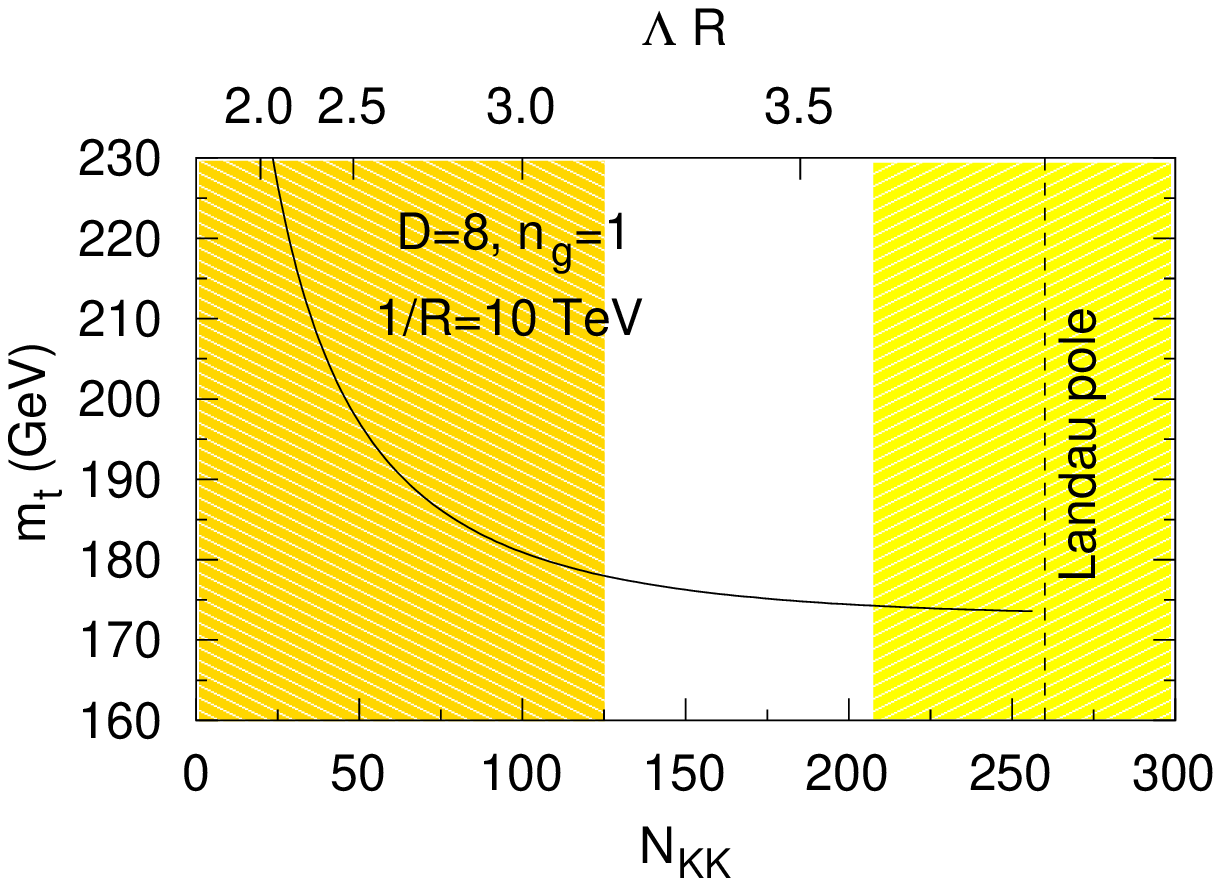}}
  \caption{Effective cutoff $\Lambda$ for the top-condensation in the bulk
           and prediction  of the top-quark mass $m_t$ (GeV) in $D=8$. 
           The unshaded regions satisfy 
           $\kappa_t (\Lambda) > \kappa_D^{\rm crit} > \kappa_\tau (\Lambda)$.
           The L.H.S. and R.H.S. graphs show behaviors of 
           $\kappa_{t,\tau}$ and $m_t$ for various cutoffs $\Lambda$, 
           respectively, where we took $1/R=10$ TeV. 
           The top and bottom lines represent
           $\Lambda R$ and $N_{\rm KK}(\Lambda)$, respectively, 
           where we used Eq.~(\ref{n_kk}) in the estimation of 
           $N_{\rm KK}(\Lambda)$. \label{fig1}}
  \end{center}
\end{figure}

Now, we are ready to compare $\kappa_t (\Lambda)$ and $\kappa_\tau (\Lambda)$
with the critical values Eq.~(\ref{kd_sd1}). 
Unless the MAC coupling exceeds at least $\kappa_D^{\rm crit}$
estimated in Eq.~(\ref{kd_sd1}),
any condensates cannot be generated in the bulk.
When Eq.~(\ref{top-cond}) is satisfied,
the top-quark acquires the large dynamical mass, whereas
the tau-lepton still remains massless.
We show our results in Fig.~\ref{fig1}. 
Actually, Eq.~(\ref{top-cond}) can be satisfied for $D=8$.
We can confirm that the top-condensation is not favored for $D=6$.
We also note that behaviors of $\kappa_{t,\tau}$
are almost unchanged for $1/R=1-100$ TeV.

\section{Predictions of $m_t$ and $m_H$}

In the same way as the approach of BHL, 
we can expect to reproduce the SM in the bulk 
in the energy scale between $1/R$ and $\Lambda$:
\begin{equation}
  {\mathcal L}_D = {\mathcal L}_{\rm kin}-y(\bar{q}_L H t_R + {\rm h.c.}) 
  + |D_M H|^2 
  - m_{H}^2 H^\dagger H 
  - \frac{\lambda}{2} (H^\dagger H)^2 \label{bulk_sm}
\end{equation}
with $M=0,1,2,3,5,\cdots, D$ and the kinetic term ${\mathcal L}_{\rm kin}$ 
for the top quark and gauge bosons.
Since we find the RGE for the top-Yukawa coupling $y$,
\begin{eqnarray}
 (4\pi)^2 \mu \frac{d y}{d \mu} &=& 
  N_{\rm KK}(\mu)\,y\,\left[\,
 \left(2^{\delta/2} \cdot N_c +\frac{3}{2}\right)y^2 
 \right. \nonumber \\ && \left. \quad 
 -C_F (6+\delta)g_3^2 - \frac{3}{4}(3-\delta/2)g_2^2 
 -\frac{(102-\delta)}{72}g_Y^2\,\right],  \label{rge_y}
\end{eqnarray}
and that for the Higgs-quartic coupling $\lambda$,
\begin{eqnarray}
 (4\pi)^2 \mu \frac{d \lambda}{d \mu} &=& 
  N_{\rm KK}(\mu)\,\left[\,
  2^{2+\delta/2} \cdot N_c \left(\lambda y^2-y^4\right) + 12\lambda^2
 \right. \nonumber \\ && \left. \quad \qquad
 +\frac{3+\delta}{4}(3 g_2^4 + 2g_2^2 g_Y^2 + g_Y^4)
 -3(3g_2^2+g_Y^2)\lambda\,\right],  \label{rge_lam}
\end{eqnarray}
we can predict $m_t$ and $m_H$ by using
compositeness conditions~\cite{Bardeen:1989ds}, 
\begin{equation}
 y(\Lambda) \to \infty, \quad 
 \frac{\lambda(\Lambda)}{y(\Lambda)^4} \to 0 .
\end{equation}

\begin{figure}[tbp]
  \begin{center}
  \resizebox{0.47\textwidth}{!}
            {\includegraphics{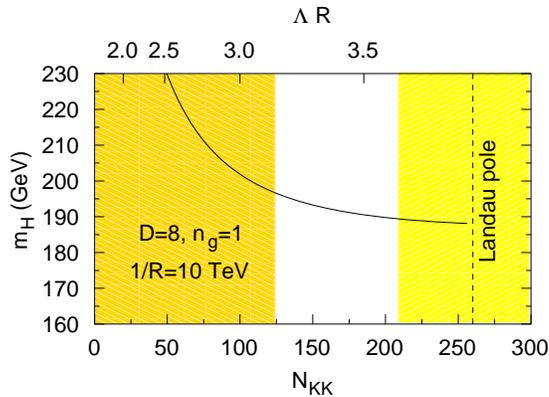}}
  \caption{Prediction of the Higgs boson mass $m_H$ for 
           $D=8, R^{-1}=10$ TeV. \label{fig6}}
  \end{center}
\end{figure}

Since the running effect of the QCD coupling $g_3$ is almost negligible
around $1/R$, the top-Yukawa coupling in $\mu \sim 1/R$ is attracted toward
the quasi IR-FP $y_*$~\cite{Hill:IR-FP},
whose behavior is approximately found as
\begin{equation}
 y_* =  g_3 \cdot \sqrt{\frac{C_F (6+\delta)}{2^{\delta/2} N_c + \frac{3}{2}}},
  \label{IR-FP}
\end{equation}
neglecting effects of the electroweak interactions.
This value obviously decreases as $\delta$ increases.
As a result, the problem of $m_t \gtrsim 200$ GeV 
in $D=4$ is suppressed in our scenario. 

Now, we predict $m_t$ and $m_H$. (See Figs.~\ref{fig1} and~\ref{fig6}.)
We obtain numerically the top-quark mass $m_t$ as
\begin{equation}
  m_t = 173-180 \; \mbox{GeV}, \label{8D-mt}
\end{equation}
and the Higgs boson mass $m_H$ as
\begin{equation}
  m_H=181-211 \; \mbox{GeV} 
\end{equation}
for $D=8$, where we took $1/R=1-100$ TeV.
For details, see Ref.~\cite{future}.

\section{Summary}

We have investigated the idea that the EWSB dynamically occurs 
due to the top-condensation in the bulk, with emphasis on 
the dynamics of bulk gauge theories.
We estimated the critical binding strength $\kappa_D^{\rm crit}$, 
based on the ladder SD equation.
We also analyzed the MAC 
by using RGEs for bulk gauge couplings. 
Combining our MAC analysis with $\kappa_D^{\rm crit}$, 
we showed that the top-condensation is favored for $D=8$,
while it is not for $D=6$.
We have predicted $m_t$ and $m_H$ in the approach of BHL.
We then obtained $m_t=173-180$ GeV, 
and $m_H = 181-211$ GeV for $D=8$.

\vspace*{4mm}

This work is supported in part by the JSPS Grant-in-Aid for the Scientific
Research (B) (2) No. 14340072.


\begin{thebibliography}{99}

\bibitem{Weinberg}
S. Weinberg, 
Phys. Rev. {\bf D13} (1976) 974; {\bf D19} (1979) 1277;
  L. Susskind, 
  Phys. Rev. {\bf D20} (1979) 2619.

\bibitem{MTY89}
  V. A. Miransky, M. Tanabashi, and K. Yamawaki, 
  Phys. Lett. {\bf B 221}, 177 (1989); 
  Mod. Phys. Lett. {\bf A 4}, 1043 (1989). 

\bibitem{Nambu89}
  Y. Nambu, Enrico Fermi Institute Report No. 89-08, 1989;
  in {\it Proceedings of the 1989 Workshop on Dynamical Symmetry
  Breaking}, edited by T. Muta and K. Yamawaki  
  (Nagoya University, Nagoya, Japan, 1990). 

\bibitem{Hill:2002ap}
  For a recent review, see C.~T.~Hill, and E.~H.~Simmons, hep-ph/0203079.

\bibitem{Bardeen:1989ds}
 W.~A.~Bardeen, C.~T.~Hill and M.~Lindner,
 Phys. Rev. {\bf D41}, 1647 (1990).

\bibitem{Hashimoto:1998tj}
  M. Hashimoto, Prog. Theor. Phys. {\bf 100}, 781 (1998).

\bibitem{Arkani-Hamed:2000hv}
 N.~Arkani-Hamed, H.~C.~Cheng, B.~A.~Dobrescu and L.~J.~Hall,
 Phys. Rev. {\bf D62}, 096006 (2000).

\bibitem{Hashimoto:2000uk}
 M.~Hashimoto, M.~Tanabashi and K.~Yamawaki,
 Phys. Rev. {\bf D64}, 056003 (2001).

\bibitem{Gusynin:2002cu}
 V.~Gusynin, M.~Hashimoto, M.~Tanabashi and K.~Yamawaki,
 Phys. Rev. {\bf D65}, 116008 (2002).

\bibitem{Raby:1979my}
 S.~Raby, S.~Dimopoulos and L.~Susskind,
 Nucl. Phys. {\bf B169}, 373 (1980).

\bibitem{future}
 M.~Hashimoto, M.~Tanabashi and K.~Yamawaki, in preparation.

\bibitem{Kobakhidze:1999ce}
 A.~B.~Kobakhidze,
 Phys. Atom. Nucl.  {\bf 64}, 941 (2001).

\bibitem{Hill:IR-FP}
  C.~T.~Hill, Phys. Rev. {\bf D24}, 691 (1981);
  C.~T.~Hill, C.~N.~Leung, and S.~Rao, Nucl. Phys. {\bf B262}, 517 (1985).

\end{thebibliography}
\end{document}